\documentclass{aa}
\usepackage{epsfig}

\newcommand{\smodels}{\mbox{cluster evolutionary synthesis models}}
\newcommand{\halfa}{\mbox{H$\alpha$}}
\newcommand{\moiii}{\mbox{$m[OIII]$}}
\newcommand{\mhalfa}{\mbox{$m(H\alpha)$}}
\newcommand{\oiii}{\mbox{[OIII]}}
\newcommand{\Msun}{\mbox{$M_{\odot}$}}
\newcommand{\Lsun}{\mbox{$L_{\odot}$}}

\newcommand{\Mcl}{\mbox{$M_{\rm cl}$}}
\newcommand{\ebv}{\mbox{$E(B-V)$}}
\newcommand{\mup}{\mbox{$M_{\rm up}$}}

\newcommand{\Mupp}{\mbox{$M_{\rm up}$}}
\newcommand{\Mmin}{\mbox{$M_{\rm min}$}}
\newcommand{\Mmax}{\mbox{$M_{\rm max}$}}
\newcommand{\Teff}{\mbox{$T_{\rm eff}$}}

\newcommand{\chinu}{\mbox{$\chi^2_{\nu}$}}
\newcommand{\chisq}{\mbox{$\chi^2$}}
\newcommand{\chir}{\mbox{$\chi_{\nu}^2$}}

\begin{document}

\title{Clusters in the inner spiral arms of M51:
   the cluster IMF and the formation history%
\thanks{
    Based on observations with the NASA/ESA Hubble Space Telescope,
    obtained at the Space Telescope Science Institute, which is
    operated by AURA, Inc., under NASA contract NAS 5-26555}
}
\author{A. Bik \inst{1,2} 
        H.J.G.L.M. Lamers \inst{1,3} 
        N. Bastian \inst{1}
        N. Panagia \inst{4,5} 
        M. Romaniello \inst{6}
        }

\offprints{H.J.G.L.M. Lamers}

\institute{ 
      {Astronomical Institute, Utrecht University, 
                 Princetonplein 5, NL-3584 CC Utrecht, 
                 the Netherlands, {\tt lamers@astro.uu.nl}}
\and  {Astronomical Institute Anton Pannekoek,
                 University of Amsterdam, Kruislaan 403,
                 NL-1098 SJ   Amsterdam, The Netherlands 
                 {\tt bik@astro.uva.nl}}
\and  {SRON Laboratory for Space Research, Sorbonnelaan 2, 
                  NL-3584 CA Utrecht, The Netherlands}
\and  {Space Telescope Science Institute, 3700 San Martin Drive,
                 Baltimore, MD21218, USA; {\tt panagia@stsci.edu}}
\and  {On assignment from the Space Science Department of ESA}
\and  {European Southern Observatory, Karl-Schwarzschild Strasse 2,
                 Garching-bei-Muenchen, D-85748, Germany; {\tt
      mromanie@eso.org}}
}
           
\date{received date; accepted date}




\abstract{
We present the results of an analysis of the \emph{HST--WFPC2} observations of
the interacting galaxy M51. From the observations in 5 broadband
filters ($UBVRI$) and two narrowband filters (\halfa\ and \oiii)
we study the cluster population in
a region of 3.2 $\times 3.2$ kpc$^2$ in the inner spiral arms of M51,
at a distance of about 1 to 3 kpc from the nucleus. 
We found 877 cluster candidates and we derived their ages, 
initial masses and extinctions by
means of a comparison between the observed spectral energy distribution
and the predictions from  cluster synthesis models for 
instantaneous star formation and solar metallicity. 
The lack of \oiii\ emission in even the youngest clusters with strong
\halfa\ emission, indicates the absence of the most massive stars and
 suggests a mass upper limit of about 25 to 30 \Msun.
The mass versus age distribution of the clusters 
shows a drastic decrease in the number of
clusters with age, much more severe than can
be expected on the basis of evolutionary fading of the clusters.
This indicates that cluster dispersion is
occurring on a timescale of 10 Myr or longer.
The cluster initial mass function has been derived from
clusters younger than 10 Myr by a linear regression fit of the 
cumulative mass distribution. This results in 
an exponent $\alpha = -d \log~ N(M) /d \log~(M) =
2.1 \pm 0.3 $ in the range of $2.5~10^3 < M < 5~10^4~ \Msun$
but with an overabundance of clusters with $M > 2~10^4$ \Msun.
In the restricted range of $2.5~10^3 < M < 2~10^4$ \Msun\ 
we find $\alpha = 2.0 \pm 0.05$. 
This exponent is very similar to the value derived for clusters in the
interacting Antennae galaxies, and to the exponent of the
mass distribution of the giant molecular clouds in our Galaxy. 
To study the possible effects of the interaction of M51 with its
companion NGC 5195 about 400 Myr ago, which triggered a huge starburst 
in the nucleus,
we determined the cluster formation rate as a function of time
for clusters with an initial mass larger than $10^4$ \Msun.
There is no evidence for a peak in the cluster formation rate
at around 200 to 400 Myr ago within 2 $\sigma$ accuracy, 
i.e. within a factor two.
The formation rate of the detected clusters decreases
strongly with age by about a factor $10^2$ between 10 Myr and 1 Gyr.
For clusters older than about 150 Myr this is due to the 
evolutionary fading of the clusters below the detection limit. For
clusters younger than 100 Myr this is due to the 
dispersion of the clusters, unless one assumes that the cluster
formation rate has been steadily increasing with time 
from 1 Gyr ago to the present time.
}
\authorrunning{A. Bik, H.J.G.L.M. Lamers, N. Bastian et al.}
\titlerunning{Clusters in the inner spiral arms of M51}
\maketitle


\section{Introduction}
\label{sec:intro}

The interaction of galaxies triggers star formation as 
clearly seen in the young starbursts in interacting 
galaxies (for reviews see Kennicutt 1998; Schweizer 1998). 
One of the best examples
is the Antennae system where the interaction is presently going on
and star formation and cluster formation occurs on a very large scale
in the region between the two merging nuclei (Whitmore et al. 1999).

The way the triggered star formation progresses through an interacting galaxy
is best studied from a slightly older interacting system, 
where the close passage of the companion is over. In such a system one
can hope to derive the age distribution of clusters as a function of
location in the galaxy and thus measure the time sequence of the 
induced star and cluster formation as a function of location in the galaxy.

One of the best systems for this purpose 
is the interacting sytem of the Whirlpool galaxy (M51)
at a distance of $8.4 \pm 0.6$ Mpc (Feldmeier et al. 1997).
The M51 system consists of a grand design spiral galaxy (NGC 5194)
interacting with its dwarf companion (NGC 5195). The almost face-on
orientation of M51 allows the observation of its structure in great
detail with  minimum  obscuration by interstellar dust. 
The interaction of these two galaxies has been modelled by various
authors, starting from the fundamental paper by Toomre \& Toomre (1972).
These authors derived an age of the closest passage to be $2 \times 10^8$
years ago.
Later refined models, e.g. by Hernquist (1990), 
and reviewed by Barnes (1998), have improved this time estimate. 
The closest approach is now
believed to have occurred about 250 - 400 Myrs ago. 
The best model is found for a distance of the pericenter of 17 to 20 kpc, 
for a mass ratio of $M/m \approx 2$ and a relative orbit which crosses
the plane of M51 under an angle of about 15 degrees. (For a discussion
of the problems with this model and possible improvements, see Barnes 1998).
Recently Salo \& Laurikainen (2000) suggested that on the basis of 
$N$-body simulations that the M51 system 
had multiple passages, with the two last at 50 -100 Myrs and
400-500 Myrs ago.

Because of its relatively small distance from us, 
and the fact that we see the M51
system face one, the system is ideally suited for the study of 
the progression of
cluster formation due to galaxy -- galaxy interaction. For this reason  we 
started a series of studies of different aspects of the M51 system
based on \emph{HST-WFPC2} observations in six broad band and two narrow band
filters.

The nucleus of M51 was studied by Scuderi et al. (2002). They found that
the core contains a starburst with an age of $410 \pm 140$ Myrs and a
total stellar mass of about $2 \times 10^7 M_{\sun}$ within the central
17 pc. This age agrees with the estimated time of closest passage of the
companion, so the starburst in the core is most likely due to the
interaction with the passing companion.
M51 contains an unresolved nucleus with a diameter smaller than 2 pc and a 
luminosity of $2\times 10^6 L_{\sun}$ (Scuderi et al. 2002).

The bulge, i.e. the reddish region with a size of 
$11 \times 16$ arcsec$^2$ = $460
\times 680$ pc$^2$, between the nucleus and the inner spiral arms, was
studied by Scuderi et al. (2002) and by Lamers et al. (2002).  
The bulge is dominated by an old stellar population with an age in
excess of 5 Gyrs. The $HST-WFPC2$ images of the bulge 
show the presence of dust lanes. The total amount of dust in the bulge
is about $2.3 \times 10^3$ \Msun\ and the dust has
about the same extinction law and approximately the same gas to dust ratio 
as our Galaxy (Lamers et al. 2002). 
This suggests a metallicity close to that in the solar neighbourhood.
The $HST-WFPC2$ images of the bulge of M51 show clearly that the dust 
is concentrated in structures in the forms of
spiral-like dust lanes and a bar that reaches all the
way down into the core. Most intriguingly is the discovery of 
about 30 bright and mainly blue point-like sources in the
bulge  that are aligned more or less along the spiral-like dust lanes.
Lamers et al. (2002) have shown that these are most likely
very young massive stars $20 < M_* < 150~ \Msun $ with little evidence
of associated clusters. This mode of star formation 
is the result of the peculiar conditions, in particular the
destruction of CO molecules, of the interstellar clouds in the bulge
of M51.

In this study we focus on the clusters 
at a distance of about 1 to 3 kpc from the nucleus, i.e. near the
inner spiral arms.
The purpose of the paper is two-fold:\\
-- (a) to determine the cluster intitial mass function, and \\
-- (b) to determine the presence or absence of a 
starburst period that can be linked to triggering by the passage of
the  companion. This is not an easy task, because our data will show
that the age distribution of the clusters is strongly affected by the
disruption of clusters older than about 40 Myrs.

We study the clusters and their properties by identifying point-like sources
in the $HST-WFPC2$ images and measuring their  
$UBVRI$ magnitudes to obtain their energy 
distributions. The magnitudes indicate that the sources are clusters
instead of single stars. 
Their energy 
distributions are compared to cluster evolutionary synthesis models to 
determine the age, mass and \ebv\ of these clusters.

In Sect. 2 we describe the observations and the data reduction. In Sect. 3 the
selection of cluster candidates is discussed. In Sect. 4 we describe
the cluster evolutionary synthesis models and in Sect. 5 the fitting
procedure for the derivation of the cluster parameters is explained.
The mass versus age distribution of the clusters is derived in Sect. 6.
 In Sect. 7 the initial mass function of the clusters is derived from the 
sample of clusters younger than 10 Myrs.
The cluster formation history is studied in Sect. 8. The summary and 
conclusions are given in Sect. 9.


\section{Observations and reduction}
\label{sec:2}

M51 was observed with \emph{HST-WFPC2} as part of the \emph{HST Supernova 
INtensive Study (SINS)} program (Millard et al. 1999). 
For this study we use the images taken in the broad band filters
F336W~($U$), F439W~($B$), F555W~($V$), F675W~($R$) 
and F814W~($I$) from the SINS program and in the narrow band 
filters F502N $(\oiii)$ and F656N ($\halfa$) from the GO-program 
of H.C. Ford. 
The image in $U$ was taken on 1994 May 12, the $BVRI$ images were
taken on Jan 15 1995 and the \oiii\ and \halfa\ images on Jan 25 1995.
The $U$ and $B$
images were split into three and two exposures of 400 $s$ and 700 $s$
respectively. 
The \oiii\ and \halfa\ images are split into two exposures of
1200 $s$ and 500 $s$ ($\oiii$), and 1400 $s$ and 400 $s$ ($\halfa$).  
In the 
remaining bands one single exposure of 600 $s$ was taken. The data was
processed through the PODPS (Post Observing Data Processing System) for
bias removal, flat fielding and dark frame correction.

To remove the cosmic rays from the $U$, $B$ \oiii\ and \halfa\ 
images, we used
the STSDAS task \emph{crrej} for combining the available exposures. 
For the $VRI$ images, where only one exposure is available, we used a
procedure called ``Cosmic Eraser''. This procedure
combines the IRAF tasks \emph{cosmicrays} and \emph{imedit} to reject as
carefully as possible the cosmic rays. The automatic detection of cosmic
rays with the task \emph{cosmicrays} is based upon two parameters, 
a detection threshold and a flux ratio. The first parameter enables 
the detection of all the pixels with a value larger than the average 
value of 
the surrounding pixels. The flux ratio is defined as the percentage of the 
average value of the four neighboring pixels (excluding the second brightest
pixel) to the flux of the brightest pixel. This parameter allows a 
classification of the detected objects: cosmic ray or star. Training objects
are used to determine the flux ratio carefully. These training objects are 
labeled by the user to be a cosmic ray or a star. With \emph{imedit} the 
detected cosmic rays signals are replaced by an interpolation of a third order 
surface fit to the surrounding pixels.

After the correction for the cosmic rays, the images were corrected for
bad pixels using the hot pixel list from the STScI $WFPC2$ website in
combination with the task \emph{warmpix}. Corrections for non-optimal
charge transfer efficiency on the CCD's of the $WFPC2$ camera were applied
using the formulae by Whitmore \& Heyer (1997).

With the task \emph{daofind} from the DAOPHOT package Stetson (1987), we
identified the point sources on the image. We performed aperture photometry
on these sources, also with the DAOPHOT package. We used an aperture
radius of 3 pixels. The sky background was calculated in an annulus with
internal and external radius of 10 and 14 pixels respectively. We only
selected the point sources with an uncertainty smaller than 0.2 in the
magnitude. Photometric zeropoints were obtained from table 28.2 of the
\emph{HST Data Handbook} (Voit 1997), using the VEGAMAG photometric
system (Holtzman et al. 1995).

The aperture correction was measured for a number of isolated, high S/N
point sources on each $WFPC2$-chip. This output was adopted for all the other
point sources on the chip. Following Holtzman et al. (1995) we have normalized
the aperture correction to 1\arcsec (10 WF pixels). The aperture
corrections we found are between -0.24 and -0.37 magnitudes. This is 
larger than the aperture corrections for stars $\approx$ -0.17 magnitudes
(Holtzman et al. 1995), which means that the detected point sources are fairly
well resolved.

We adopt a distance of $d = 8.4 \pm 0.6$ Mpc (Feldmeier et al.
1997), which
corresponds to a distance modulus of 29.62. At this distance, 1 \arcsec\
corresponds to a linear distance of 40.7 pc, which means that an 
$HST-WFC$ pixel of 0.1\arcsec\ corresponds to 4.1 pc.


\section{Selection of the clusters}
\label{sec:3}

\begin{figure*}
\epsfig{figure=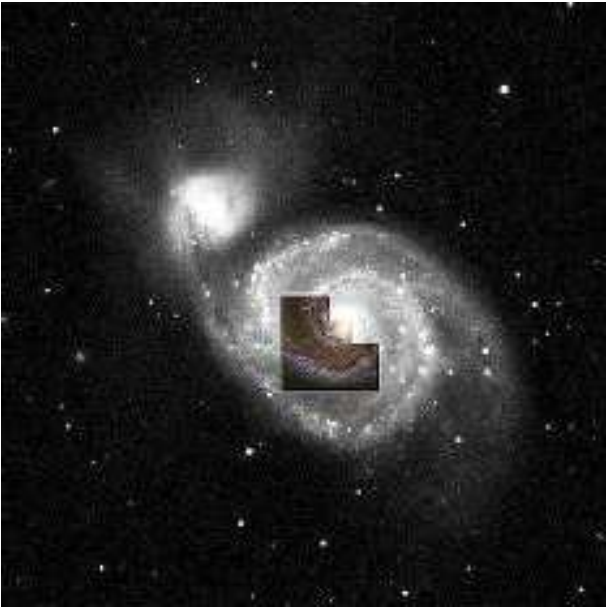, width=8.0cm}
\hfill
\epsfig{figure=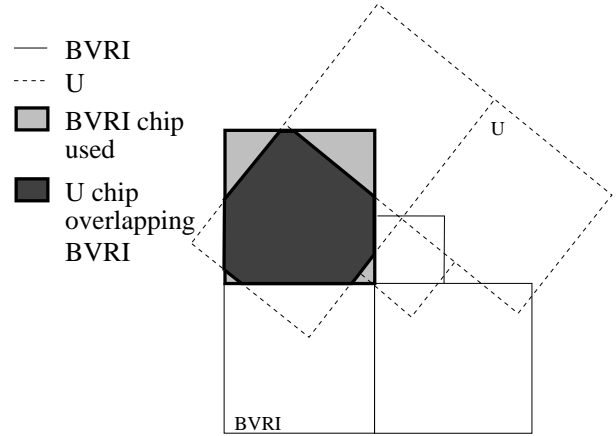,width=8.0cm}
\caption[]{The location of the $WFPC2$ camera when the $BVRI$
  images were taken, superimposed on an optical image of M51.
Right: The orientation of the $WFPC2$ camera when the $BVRI$ and the
$U$ images were taken. The grey area indicates the location of the
$WF2$ chip, covering  an area of $3.25 \times 3.25$ kpc (for an adopted
distance of 8.4 Mpc), 
that was used in this analysis. Its exact location and orientation are
given in Fig. 2. The dark grey region
shows the overlap of the $BVRI$ and $U$ images. }
\label{fig:orientation}
\end{figure*}

The $U$-band images were taken in 1994, 8 months earlier than the images in the
other passbands. The orientation of $HST-WFPC2$ was not the same at
the two epochs. In the 1995 $BVRI$-images the nucleus of M51 was in
the center of the PC-image.  The $U$-image was centered on SN1994I
(that was 15'' off from the nucleus) so that the nucleus was near the
edge of the PC-image. The orientation of the $WFPC2$-images is shown
in Fig. \ref{fig:orientation}.

For our analysis only the part of $WF4$-chip of the $U$
image
that overlap with the $WF2$-chip of the $BVRI$ images is used (see Fig. 1).
To identify the various sources, we obtained a position of
the sources in $U$ and transformed these positions to the
coordinate frame of the $BVRI$ images. 
The  orientation of the \oiii\ and \halfa\ images is very similar to
those of the $BVRI$ images. We transformed these narrow band images
to the same orientation as the $BVRI$ images.
 
We applied our analysis of the cluster energy distributions 
to the sources that were detected in the images of 
at least three of the $BVRI$ broad-band filters with a magnitude uncertainty
smaller than 0.2 in each band. 
To this purpose we compared the positions of all detected sources
in the $BVRI$ images. If objects occur in at least three images
within a position tolerance of 2 pixels they are selected for our study.
This tolerance was chosen on the basis of tests with 
clearly identifiable sources that were observed in most bands.
For the sources detected in at least three images 
we then checked for the presence of a source in the $U$ image,
in the region where the $U$ image and the $BVRI$
images do overlap (the dark grey region in Fig. 1). 
If a source was not detected in a particular band, we adopted a
conservative lower limit for the magnitude in that band (see below).
This was not always possible for the $U$-magnitude, because many of
the sources detected in the $BVRI$ bands are located in an area that was
not covered by the $U$-image. As a consequence, we have no information
on the $U$-magnitude for many of the sources.
 
\begin{figure}
\epsfig{figure=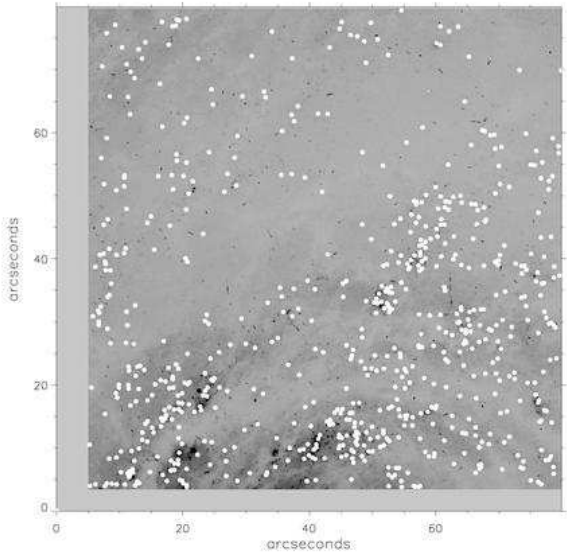,width=8.0cm}
\caption[]{Left: the location of all the detected point sources in the
  WF2 chip superimposed on the V-band image. 
  This image is rotated 90 degrees clockwise compared to 
  Fig. 1.
  The coordinates of the lower-left and upper-right 
  corners of the image are:
  pixel(1,1):
  RA(2000)=13$^h$29$^m$55$^s$.90, Dec(2000)=47$^0$11'27''.2. and
  pixel{800,800}:
  RA(2000)=13$^h$29$^m$57$^s$.94, Dec(2000)=47$^0$13'16''.6. }
\label{fig:sources}
\end{figure}

The location of
the sources in the $WF2$-chip is shown in Fig. \ref{fig:sources}.
The overlay of the sources on the $V$ band image  clearly
shows that the clusters are concentrated in or near the spiral arms
at a distance of about 1.5 kpc from the nucleus of M51.
The number of sources, detected in the various bands, 
is listed in Table \ref{tbl:1}. The sample contains a total of
877 sources for which we have reliable photometry ($\sigma < 0.20$
magn.) in at least three bands. 
\footnote{The coordinates and the photometry of the objects 
are available in the electronic version of A\& A}

\begin{figure}
\centerline{\epsfig{figure=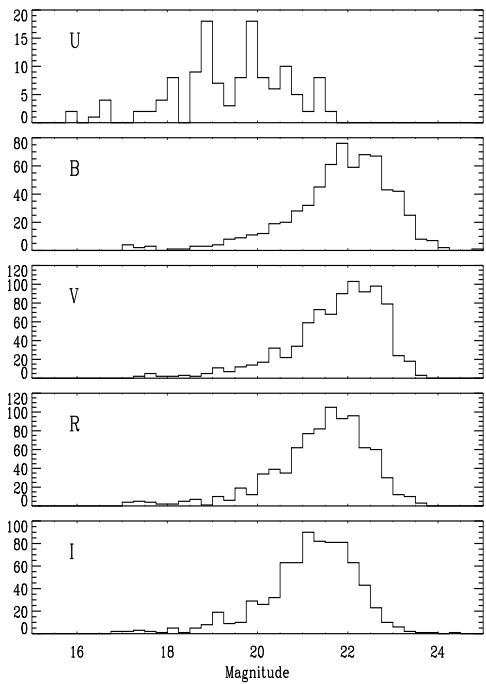,width=10.0cm,height=10.0cm}}
\caption[]{The histogram of the magnitudes in the different bands.
The slow increase at bright magnitudes is due to the
luminosity function and the steep decrease at fainter
magnitudes is due to the detection limit.}
\label{fig:hist}
\end{figure}

Fig. \ref{fig:hist} shows the magnitude distribution of the sources
in the various bands. 
All distributions show a slow increase in numbers towards fainter
objects, a maximum and a steep decrease towards even fainter
sources. The slow increase to fainter sources reflects  their 
luminosity function.
The steep decrease to high magnitudes is due to the detection limit.
This detection limit is not a single value for each band, because
of the variable background in the $WF2$ field due to the spiral arms.
The maximum of the distributions are near $U \simeq 20.0$, 
$B \simeq 22.0$, $V\simeq 22.0$, $R \simeq 21.5$ and $I \simeq 21.0$.
These values will be adopted as conservative 
lower limits for the magnitudes of sources
that were not detected in any given band.

All the objects fall in the  visual magnitude range of  
$16.5 <V < 23.6$. For
distance modulus of 29.62 this corresponds to an absolute magnitude
range of 
$-12.1 < M_V <-6.0$ in the case of zero extinction and $-12.6 < M_V<-6.5$
in case of moderate extinction with $E(B-V) \simeq 0.17$. This means that the 
faintest objects
could be either very bright stars or small clusters. The vast
majority of the selected point sources are so bright that they must be
clusters. 
(See Sect \ref{sec:5.5}).

\begin{table}
\caption[]{Numbers of point sources detected in the different
$WFPC2-HST$-images}
\begin{tabular}{cccccr}
\hline\hline\noalign{\smallskip}
F336W  & F439W & F555W & F675W & F814W & number \\
$(U)$ & $(B)$ & $(V)$ & $(R)$ & $(I)$ & objects \\
\hline
\noalign{\smallskip}
d&d&d&d&d&  76 \\
l&d&d&d&d& 366 \\
n&d&d&d&d& 109 \\
d&l&d&d&d&  30 \\
l&l&d&d&d& 144 \\
n&l&d&d&d&  39 \\
d&d&d&d&l&  21 \\
l&d&d&d&l&  78 \\
n&d&d&d&l&  14 \\
\hline 
\noalign{\smallskip}
 & & & & & 877 \\
\hline
\end{tabular}\\
$d$  means "detected", $l$ means "lower magnitude limit" (i.e. the
object is fainter than the detection limit) and $n$ means
"not observed".
\label{tbl:1}
\end{table}



\subsection{The \halfa\ and [OIII] magnitudes}
\label{sec:3.1}

The magnitudes in the narrow-band images of filters $F656N$ (\mhalfa )
and $F502N$ (\moiii )  can be compared with those of the wide-band
filters   $F675W$ ($R$) and $F555W$ ($V$) respectively to derive a
measure of the equivalent width  of the \halfa\ and \oiii\ lines.  When
the magnitudes are expressed in the ST-system, the magnitude
differences $R-\mhalfa$ and $V-\moiii$ are by definition equal to 0.0
if the spectrum of a source is a featureless, flat  continuum 
(i.e.  the spectrum of the source is a continuum $F_\lambda=const$
without a photospheric \halfa\ absorption, and there is   no HII region
around the star). The differences $R-\mhalfa$ and $V-\moiii$
are either positive or negative if the source
spectrum has an emission or an absorption line, respectively, falling
in the  narrow band filter, independently of the interstellar
extinction.  With the Vega-system magnitudes one has to fold in both
the non-zero spectral slope and the
discrete features present in $\alpha$ Lyrae's spectrum. Therefore,  a
featureless flat continuum would correspond to $R-\mhalfa \simeq +0.08$ and
$V-\moiii \simeq -0.26$ colors in the Vega-system.

\begin{figure}
\centerline{\epsfig{figure=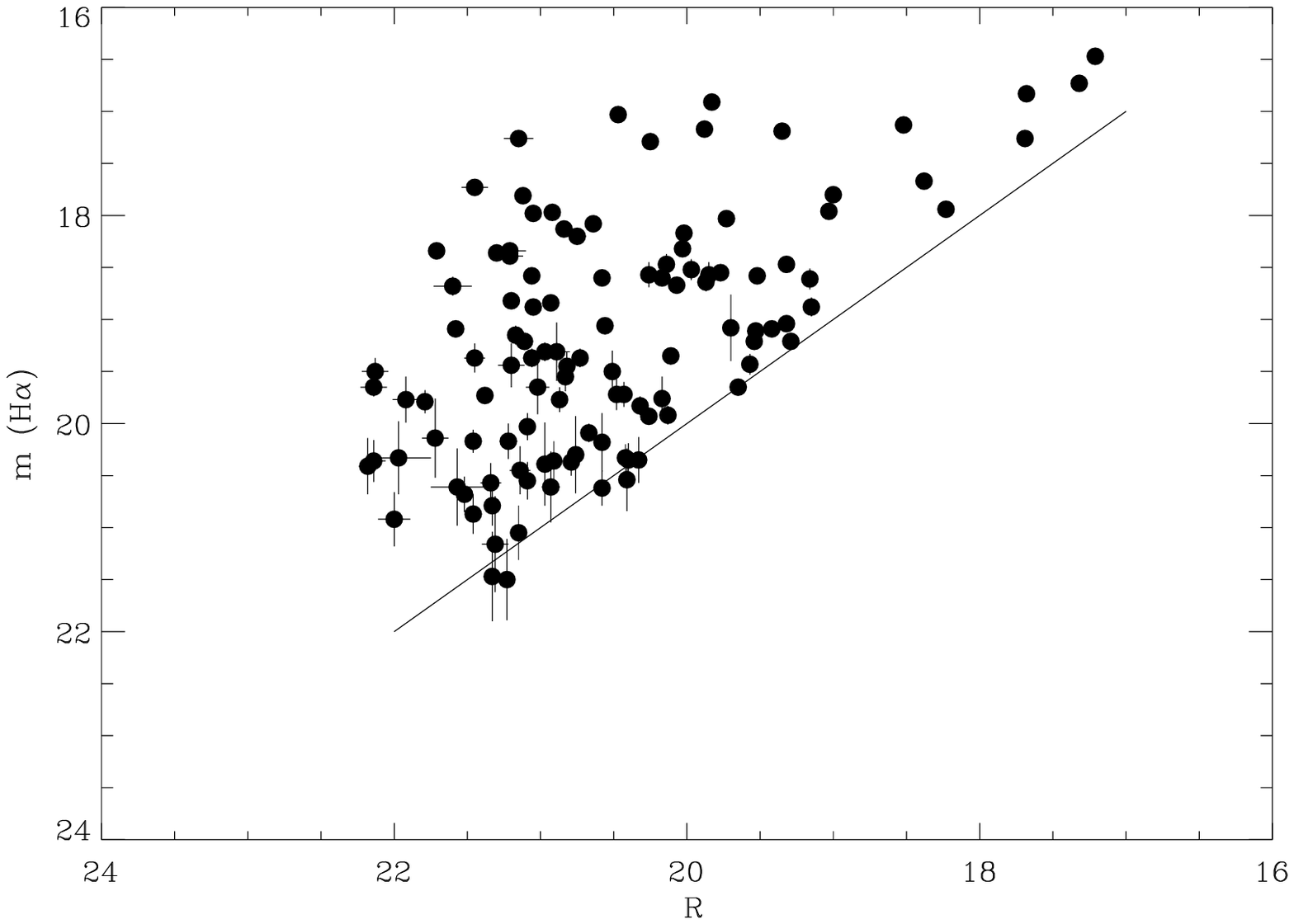,width=8.4cm}}
\centerline{\epsfig{figure=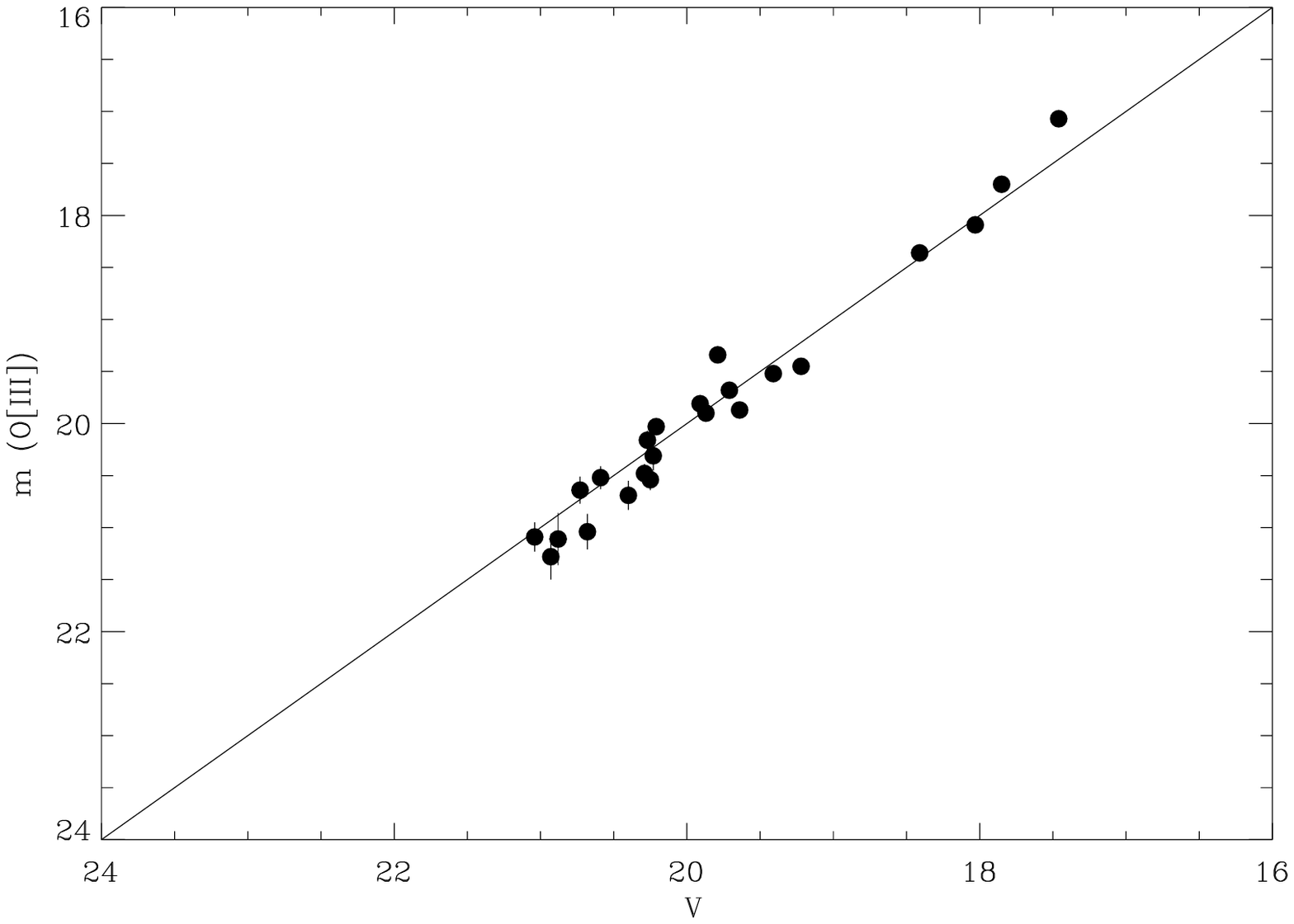,width=8.4cm}}
\caption[]{The top figure shows \mhalfa\ versus $R$
and the lower figure shows \moiii\ versus $V$ for objects were
these magnitudes could be determined with an accuracy better than
0.2 magn. Notice that many objects have \halfa\ emission, but no 
object has detectable \oiii\ emission. }
\label{fig:halfa}
\end{figure}

 Fig. \ref{fig:halfa}a shows the \mhalfa\ versus $R$ and  Fig.
\ref{fig:halfa}b shows \moiii\ versus $V$ for the subset of sources for
which we could measure these magnitudes with an accuracy better than
0.2. The figure shows that many sources have \halfa\ emission ($\mhalfa
< R$) but none (!) of the sources has \oiii\ emission (note that
\moiii\ is slightly fainter than the  $V$ magnitude as appropriate for
complete absence of an emission line).  The lack of \oiii\ emission
indicates that the far-UV flux, at $\lambda < 350$ \AA\, i.e. at
energies higher than the second ionization potential of oxygen, is
quite low. For main sequence stars with  $L>2.5 \times 10^5 \Lsun$
(i.e. with $M > 30 \Msun$ and  $\Teff > 33~000$ K), surrounded by an
HII region with approximately solar abundances, we would expect $ V -
\moiii > R - \mhalfa$. We will show later, on the basis of the study of
the energy distributions, that many of the objects  are very young
clusters with ages less than a 10 Myr. The lack of \oiii\ emission
shows that these clusters do not contain stars with \Teff\ above about
30~000 K, which corresponds to about 25 to 30 \Msun\ (e.g. 
Chiosi \& Maeder 1986). We can exclude the possibility that the lack of
\oiii\ emission be due to just a high metallicity, which could reduce
the electron temperature in an HII region and weaken optical forbidden
lines considerably,   without requiring a lower effective temperature,
or, equivalently, a low  upper mass  cutoff. This is because 
in many HII regions in M51 an in other metal-rich spiral galaxies
not only are the
[OIII] lines faint or absent but also the HeI lines are unusually faint
relative to Balmer lines (Panagia 2000; Lenzuni \& Panagia, in
preparation).  This result indicates that He is only
partially ionized and, therefore, that the ionizing  radiation field is
indeed produced exclusively by  stars with effective temperatures much
lower than 33,000K .
So we conclude that the upper mass limit for clusters in the inner
spiral arms of M51 is about 25 to 30 \Msun.

\section{Cluster evolution models}
\label{sec:4}

To determine the age, initial mass and  $E(B-V)$ of the clusters, 
we compare the observed energy distributions with energy distributions derived
from theoretical \smodels. We used two sets of models: the Starburst99
models for ages up to 1 Gyr, and the Frascati models for ages of 10
Myr to 5 Gyr.
In this section the models are described.

\subsection{Starburst99 cluster evolution models}
\label{sec:4.1}

To determine the age, initial mass and  $E(B-V)$ of the clusters,  we
compare the observed energy distributions with energy distributions
derived from theoretical \smodels. We used two sets of models: the
Starburst99 models for ages up to 1 Gyr, and the Frascati models for
ages of 10 Myr to 5 Gyr. In this section the models are described.

\subsection{Starburst99 cluster evolution models}
\label{sec:4.1}

We compared the observed spectral energy distributions (SED) of our
objects with the \smodels\ of Leitherer et al. (1999), i.e. the 
Starburst99 models for instantaneous star formation. In these
models, the stellar atmosphere models of Lejeune et al. 
(1997) are included. For
stars with a strong mass loss the Schmutz et al. (1992) extended model
atmospheres are used. The stellar evolution models of the Geneva
group are included in these models.  For our analysis we used the models of
the Spectral Energy Distributions (SEDs).

In the cluster models the stars are assumed to be formed with  a
classical Salpeter IMF-slope of $d ~\log (N) /d~\log (M) =-2.35$. The
lower cut-off mass is $M_{\rm low}$ = 1~M$_{\sun}$. We adopt two values
for the upper cut-off mass: the standard value of $M_{\rm up}$ =
100~M$_{\sun}$, and the value of $\mup = 30~ \Msun$.  This last value
is adopted because it is suggested by the lack of \oiii\ emission (see
\ref{sec:3.1}). The total initial mass of all models is $10^6$ \Msun.
Leitherer et al. (1999) present models for 5 different metallicities,
from $Z= 0.001$ to $Z=0.040 = 2 \times Z_{\sun}$. Observations of
HII-regions have shown that the metallicity of the inner region of M51
is approximately $2  Z_\odot$  or slightly higher (e.g. Diaz et
al. 1991; Hill et al. 1997). Therefore we adopt the models with 
$Z=0.020$ and 0.040. The almost solar metallicity agrees with the study
of Lamers et al. (2002), who found that the extinction properties  and
the gas to dust-ratio  in the bulge of M51 is very similar to that in
the Milky Way.  So we use four sets of models, denoted by the pair
$(Z,\Mupp)$ of (0.02,100), (0.02,30), (0.04,100) and (0.04,30).  For
these combinations we calculated the SEDs of the cluster models, 
taking into account the nebular continuum emission. 

To obtain the  absolute magnitudes from the  theoretical spectral energy
distributions  in the five broad-band filters and the two narrow-band filters 
we convolved the predicted spectra of the Starburst99 models with the $WFPC2$ 
filter profiles.  
The spectrum of Vega
was also convolved with the filter functions in order to find the
zero-points of the magnitudes in the VEGAMAG system.
For each combination of $Z$ and $\Mupp$ we calculated the SED of 195 models
in the range of 0.1 Myr to 1 Gyr, with time steps increasing from 
0.5 Myr for the youngest models to 10 Myr for the oldest models.

The adopted lower mass limit of 1 \Msun\ for the cluster stars has
consequences for the derived masses of the clusters. If the IMF with a
slope of -2.35 has a lower mass limit of 0.2 \Msun, as for the
Orion Nebula Cluster (Hillenbrand 1997), the derived masses of the
clusters would be a factor 2.09 higher for an upper limit of 30
\Msun. If the lower limit is 
0.6 \Msun, the mass of the clusters will be a factor 1.28
higher than those of the Starburst99 models for $M_{up}=30$ \Msun. 
We will take this effect into account in the determination of the cluster masses.

\begin{figure}
    \centerline{\epsfig{figure=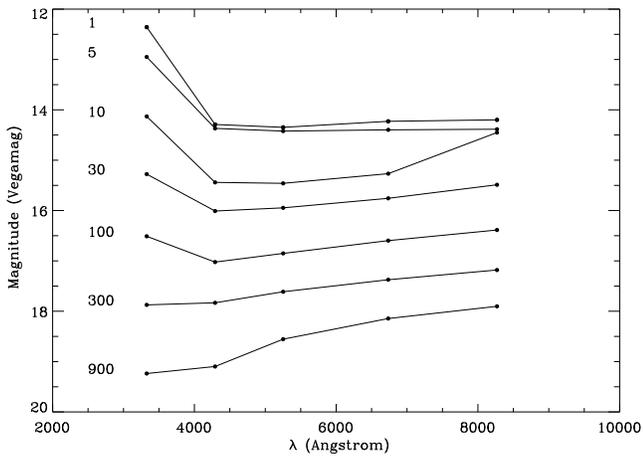,width=9.0cm}}
    \caption[]{Examples of some characteristics of the theoretical energy
     distributions, in magnitude versus wavelength (\AA) of the
     $UBVRI$ filters (from left to right), 
     for STARBURST99 clusters with an initial mass of 
     $10^6$ \Msun\ at the distance of M51.
     The age of the cluster in Myrs is indicated.  
     The evolutionary variation is discussed in Sect. 4.1}
   \label{fig:starburst99}
\end{figure}

Some of the theoretical energy distributions in the $WFPC2$
wide-band filters used in this study are shown in
Fig.~\ref{fig:starburst99}, for clusters with an initial 
mass of $10^6$ \Msun\ at the distance of M51.
During the first 5 Myrs the UV flux remains high because most of 
the O-type stars have main sequence ages longer than 5 Myr.
Between 5 and 10 Myr the O-type stars disappear. This results in a
general decrease at all magnitudes, except in the $I$ band 
where the red supergiants start to dominate. At later ages the flux
decreases at all wavelengths as stars end their lives. After 900 Myrs
only stars of types late-B and later have survived.

Fig. \ref{fig:starburst99} shows that at any wavelength in the
observed range the flux decreases with time. 
This fading of the clusters indicates that the detection limit of the
clusters will gradually shift to those of higher mass as the clusters age.
This is because the flux of a cluster (for a given
stellar initial mass function) at any age and wavelength
is proportional to the initial number of stars. The more massive
the initial cluster, the brighter the cluster and the longer
the flux will remain above the detection limit as the cluster fades
due to stellar evolution. For any given detection limit, we can
calculate the lower mass limit of the observable cluster as a function of age.

\begin{figure}
\centerline{\epsfig{figure=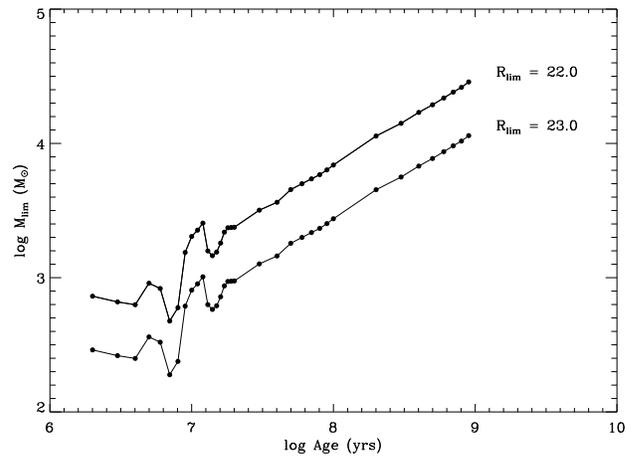,width=9.0cm}}
\caption[]{The relation between the critical initial 
mass $M_{\rm lim}$ and age, derived for two values of the  
$R$-band magnitude limit; 22.0 and 23.0 magn. It is calculated from
the Starburst99 cluster models, for a distance of 8.4 Mpc and 
no extinction. Only clusters with an initial  mass higher
than $M_{\rm lim}$ can be detected as faint as $R_{\rm lim}$
at time $t$. Alternatively,
clusters with a mass $M_{\rm lim}$ can only be detected 
down to a magnitude $R_{\rm lim}$
when they are younger
than the age $t$, indicated by these relations.
}
\label{fig:rlim}
\end{figure}

Fig. \ref{fig:rlim} shows the relation between the initial mass of
a cluster and its age when it fades below the detection limit, derived
from the Starburst99 models. The $R$-magnitude ($R$) of a cluster
of initial mass $M_i$ 
without extinction at the adopted distance of 8.4 Mpc of M51 ($dm=29.62$)
is related to the absolute $R$-magnitude ($M_R$) predicted for the
Starburst99 models of $10^6$ \Msun\ by

\begin{equation}
R(t)~=~M_R(t) + 29.62 - 2.5 \times \log (M/10^6)
\label{eq:Rmag}
\end{equation}
So, for a given detection limit $R_{\rm lim}$ the critical mass 
$M_{\rm lim}$ of a
cluster that can be detected at age $t$ is given by

\begin{equation}
\log M_{\rm lim}(t)~=~6+0.4 \times(M_R(t)+29.62-R_{\rm lim})
\label{eq:mlim}
\end{equation}
We have adopted the $R$ magnitudes to calculate the observed lower
limits of the distributions for two reasons: (a) all clusters
were detected in the $R$ band and (b) the effect of extinction is
small in the $R$ band.
The resulting relation between $M_{\rm lim}$ and age is shown in Fig.
\ref{fig:rlim} for detection limits of $R_{\rm lim}$= 22.0 and
23.0 magn. We will use these
limits later, in Sect. 5, to explain the distribution of 
clusters in the mass versus age diagram.

\begin{figure}
\centerline{\epsfig{figure=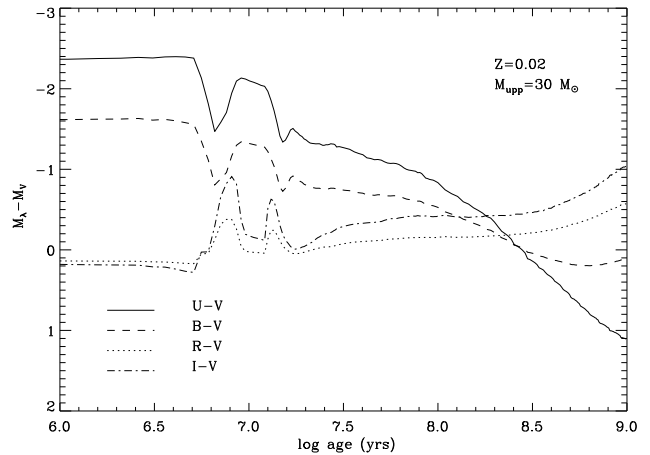,width=9.0cm}}
\caption[]{The variations of colours with age of the Starburst99 models
of $Z=0.02$ and an upper mass limit of 30 \Msun.
Notice the periods of rapid variability near the peaks and dips around
$t \simeq 6$ and $16$ Myrs, $\log (t) \simeq 6.8$ and 7.2. }
\label{fig:sb99colours}
\end{figure}

Fig. \ref{fig:sb99colours} shows the variations of the colours
of the Starburst99 models for solar metallicity and with an upper mass
limit of 30 \Msun. The figure shows that the colours do not change 
monotonically, but that there are peaks and dips, especially in the
age range of $6.5 < \log(t) < 7.3$. These are related to the phases of
the appearance of red supergiants that also produced the dips in the
limiting magnitude curves of Fig. \ref{fig:rlim}.

\subsection{The Frascati models}
\label{sec:4.2}

Since the Starburst99 models only cover the age range up to 1 Gyr,  we
also used synthetic cluster models for older ages from the Frascati group.
These ''Frascati-models'' were calculated by Romaniello
(1998) from the evolutionary tracks of Brocato \& Castellani (1993) and
Cassisi et al. (1994) using the $HST-WFPC2$ magnitudes derived from the
stellar atmosphere models by Kurucz (1993). 
These models are for instantaneous formation
of a cluster of solar metallicity stars in the mass range of 0.6 to 25
\Msun, with a total initial mass of 50~000 \Msun,
distributed according to Salpeter's IMF. These models cover an
age range of 10 to 5000 Myr. There are 48 models with timesteps
increasing from 10 Myr for the youngest ones to 500 Myr for the oldest
ones. We have increased the magnitudes of the Frascati models 
by -3.526 magn. in order to scale these models to a total mass
of $10^6$ \Msun\ in the mass range of 1.0 to 25 \Msun; i.e. $-$3.25 magn.
correction for the conversion from $5~10^4$ to $1~10^6$ \Msun, and
0.279 magn. correction for the conversion from a lower stellar mass limit of
0.6 \Msun\ to 1.0 \Msun.
In this way
the masses can be compared directly with those of the Starburst99 models.

The Starburst99 models are expected to more accurate for the younger
clusters, because they are based on the evolutionary tracks of the
Geneva-group which include massive stars. The Frascati models are
expected to be more accurate for the old clusters, because they
include stars with masses down to 0.6 \Msun.


\subsection{The extinction curve}
\label{sec:4.3}

To determine the values of $E(B-V)$ of the clusters we assume the mean 
galactic reddening law.
Lamers et al. (2002) and Scuderi et al. (2002) 
have shown that the extinction law of respectively the bulge and 
the core of M51 agree very well with the galactic law. 
The values of $R_i=A_i/E(B-V)$ for the $HST-WFPC2$ filters
have been calculated by Romaniello (1998). They are
respectively 4.97, 4.13, 3.11, 2.41 and 1.91 for the 
$U, B, V, R$ and $I$ bands.


In the fitting of the energy distributions we did not take into
account the disruption of clusters. We assumed that the stars 
contribute to the total energy distribution of the clusters until
the cluster is completely dissolved and is no longer detectable
as a point source. This is a reasonable assumption
because we measured the magnitudes in a circle of 40.7 pc radius
(see Sect. 2).


\section{Fitting the observed energy distributions to observed cluster models}
\label{sec:5}

\subsection{The fitting procedure}
\label{sec:5.1}

We have fitted the energy distributions of the clusters with good photometry in
four or more filters, 
 viz. $UBVRI$, $BVRI$, $UVRI$ and $UBVR$ (see Table \ref{tbl:1}),
with the energy distributions of the 
\smodels, discussed above,  using a three dimensional maximum 
likelihood  method. The three fitting parameters are:
 the age of the cluster ($\tau$), the reddening ($E(B-V)$) and 
the initial mass of the cluster ($M$). For the clusters detected 
in only three filters
we reduce the parameter space and make a two dimensional
maximum likelihood fit. For objects not detected in a band, we
adopted the magnitude lower limits (described in Sect. \ref{sec:3})
to check which fits were acceptable.\footnote{We will refer to this method
in subsequent papers as the "3(2)DEF- method'', i.e. the three (or
two) dimensional energy distribution fitting''.}

\subsection{Three dimensional maximum likelihood method}
\label{sec:5.2}

This method was applied to sources that are detected in at least
four bands.
We fitted the observed energy distributions with those predicted for the
Starburst99 cluster models and the Frascati models
 (see Sect. \ref{sec:4.1}). For each model-age we have two fitting
parameters $M$ and \ebv. The cluster models are for an initial cluster
mass of $10^6$ \Msun. For other masses the flux simply scales 
$M/10^6 \Msun$. We have adopted an uncertainty in the model
fluxes of 5 percent (0.05 magnitudes) in all bands.

To reduce the range in masses in the parameter space, we make an
initial guess of the cluster mass based on the observed magnitudes,
since the distance to M51 is known. 
By calculating the average difference in magnitude -- weighted by the errors --
between the theoretical energy distribution for $10^6$ \Msun\
and the observed one, we have a 
good first approximation of the mass of the cluster:

\begin{eqnarray}
\label{eq:mass}
& M_{guess}(E(B-V),\tau) \simeq  \nonumber \\
& \frac{\sum_{\lambda} 
1/\sigma_{\lambda}\{m_{\lambda}^{obs} -
m_{\lambda}^{mod}(E(B-V),\tau)\}}{\sum_{\lambda}1/\sigma_{\lambda}}.
\end{eqnarray}
with \ebv\ between 0.0 and 2.0 in steps of 0.02, where $\sigma$ is the
uncertainty in the magnitude.
This is a reasonable range, compared to the average colour excess for 
the bulge of 0.2 found by Lamers et al. (2002).
With this initial guess we then make a three dimensional likelihood
analysis, using a mass range from 1/1.5 to 1.5 times the initial mass
estimate in steps of 0.004 dex. The extinction was varied between 0.0
and 2.0 in steps of 0.01 and the age was varied between 0.1 Myr and 1
Gyr for the Starburst99 models and between 10 Myr to 5 Gyr for the
Frascati models.

\begin{figure*}
    \centerline{\epsfig{figure=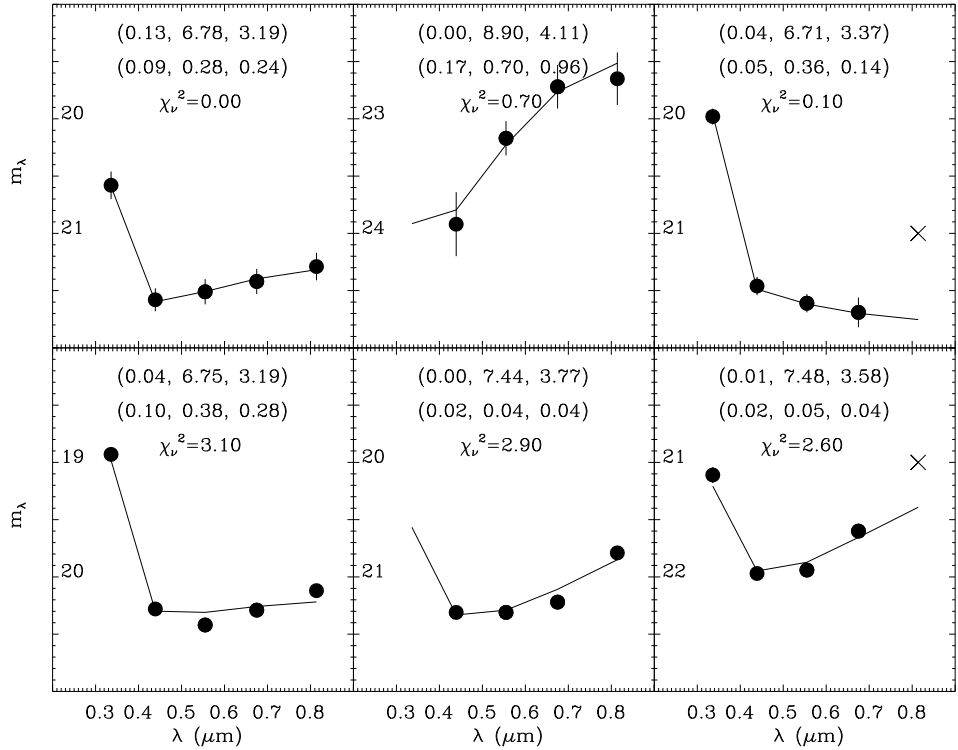,width=16.0cm}}
    \caption[]{Example of some fits. 
    The filled dots represent the observed energy
    distributions of objects measured in at least four bands,
    with their error bars in magnitude versus
    wavelength (in $\mu m$). In several cases the error bars are
    smaller than the size of the dots.     
    Crosses indicate magnitude lower limits.
    The parameters of the fits are
    indicated in an array $(E(B-V),~\log(t),~\log(\Mcl))$.  
    The second
    array gives the uncertainties in these parameters.
    The line is the best fit obtained with the maximum likelihood method. 
    Top figures: fits with $\chinu < 1$; bottom figures: fits with
    $\chinu \simeq 3$. Notice that the uncertainty of the fit
    parameters depends not only on $\chi_{\nu}^2$ but also 
    strongly on the quality of the data. 
    For instance, the errors
    in the magnitudes of the last two clusters are very small, so that
    the fit parameters are accurate, despite the high
    value of $\chi_{\nu}^2$.
}
    \label{fig:fits}
\end{figure*}

For every fit we obtain a value for the reduced $\chi^2$, i.e.
\chinu\ = \chisq/$\nu$, where
$\nu$ is the number of free parameters i.e. the number of the observed
data points minus the number of parameters in the theoretical model.
 For a good fit, \chinu\ should be about unity. 
We checked that the fits were consistent with the faint magnitude limits
of the filters in which the object was not detected. If not, the fit was
rejected. The fit with the minimum value of $\chi^2_{\nu}$
was adopted as the best fit and the corresponding values of
\ebv, age and $M_i$ were adopted.
This method was applied for fits with the four sets of the Starburst99
models and with the Frascati models.
Fig. \ref{fig:fits} shows some examples
of the results of the fitting process.

To estimate the uncertainty in the determined parameters we use
confidence limits.
If $\chinu < \chinu (min)+1$ then the resulting parameters, i.e. $\log (t)$,
$\log (M_i)$ and \ebv, are within the 68.3 \% probability range.
 So the accepted ranges in age, mass and extinction are derived from the
fits which have $\chinu (min) < \chinu < \chinu (min)+1$.
With this method we derived the ages, initial
masses\footnote{The initial cluster mass that we derived in
  this way is in fact ``the initial mass of the current stellar
  population'', i.e. it is the ``current mass corrected 
  for stellar evolution effects''.
  This value can be different from the 
  the initial mass of the original cluster if the cluster suffered
  evaporation or disruption. In that case the derived initial mass is the 
  initial mass of the original cluster minus the fraction that has
  disappeared by evaporation or disruption.}
and extinction with their uncertainties of 602 clusters.

\begin{figure}
    \centerline{\epsfig{figure=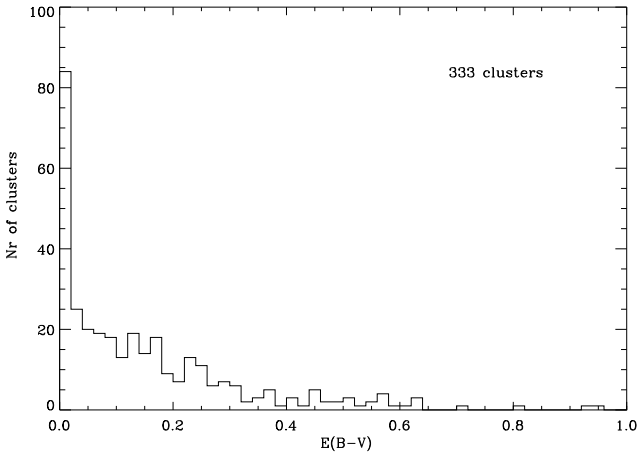,width=9.0cm}}
    \caption[]{Histogram of the values of \ebv\ from the fits with 
    observations in
    $BVRI$ and $UBVRI$. The extinction is small: 90\% of the clusters
    have $\ebv<0.40$}
    \label{fig:ebvhist}
\end{figure}

Fig. \ref{fig:ebvhist} shows a histogram of number of clusters 
as function of 
\ebv. We will use this distribution for the two dimensional maximum
likelihood fit of clusters detected in three bands only.
 The redding is small: 90 \% of the clusters have a reddening 
lower than 0.40; 67 \% of the clusters have a reddening 
lower than 0.18 and 23\%  have no detectable reddening.

\subsection{Two dimensional maximum likelihood method}
\label{sec:5.3}

For the clusters observed in only three bands, it is not possible to make
 a three dimensional maximum likelihood fit. 
To reduce the parameter space we adopted the probability distribution
of \ebv\ in the range of \ebv\
 between 0.0 and 0.4, as shown in Fig. \ref{fig:ebvhist}. 
For every value of \ebv\ between 0.0 and 0.40 (in steps of 0.02) 
and for every age of the model cluster, we first determine the 
mass of the cluster
by means of Eq. (\ref{eq:mass}). The results of the three dimensional 
fits have shown this is a good approximation. This results in a 
maximum likelihood age of the cluster for every value of \ebv.

 The \chinu\ which comes out from the two dimensional maximum
likelihood method is used to distinguish between the accepted and rejected
fits. For the parameter $\nu$ we used a value of 3-1=2, because we only fit 
the age of the cluster.  
The mass of the cluster comes from the scaling
of the magnitudes to the best-fit model.
A fit is accepted if it  agrees with
the lower magnitude limits in the filters where it was not measured.
To determine the age of the cluster, we average the
ages, weighted by the probability that each particular value of \ebv\
occurs, derived by normalizing the distribution in
Fig. \ref{fig:ebvhist} in the range of $0.0 < E(B-V)<0.40$.
The error in the age is determined by calculating the value of 
the standard deviation $\sigma$ of the age, again weighted
with the probability that the value of \ebv\  occurs. 
The value of \ebv\ is determined by using the one which belongs to the
model with the age closest to the average age. The initial
mass of the cluster is then calculated by Eq. (\ref{eq:mass}) for the
adopted value of $E(B-V)$. With this method we derived the age,
initial mass and the extinction of 275 clusters, using both the
Starburst99 models and the Frascati models.

\subsection{Starburst99 or Frascati models?}
\label{sec:5.4}

We have compared the energy distributions of the objects with five
sets of models: four sets of Starburst99 models, 
with $Z=Z_{\odot}$ or $2Z_{\odot}$
and $\Mupp=30$ or $100$ \Msun, and the Frascati models with $Z=Z_{\odot}$
and $\Mupp=25$ \Msun.
Based on the lack of \oiii\ emission we adopt the models with $\Mupp =
25$ or 30 \Msun. The models with $\Mupp=100$ \Msun\ are only used 
later to check the influence of this adopted upper limit on our results.

We have compared the ages derived from fitting the observed SEDs with
the Starburst99 models and the Frascati models, for objects measured
in at least four filters and with fits of $\chinu < 3.0$. 
We found that in the age range of
10 to 800 Myr there is a reasonable correlation between the results of
the SB99 models and the Frascati models, with the Frascati models
giving ages systematically about 0.4 dex smaller than the SB99 models.
In this age range we adopt the SB99 models, because they are based on
more reliable evolutionary tracks, better stellar atmosphere
models and because the nebular continuum is included. We find that
in this mass range the fits with the SB99 models have smaller 
$\chir$ than the fits with the Frascati-models. 
For objects with ages above about 700 Myrs ($\log (t)
 > 8.85$) the Frascati-models give the most reliable fits with
smaller $\chir$ than the SB99 models.
This is probably because the lower mass
limit of the SB99 models is 1 \Msun, and the Frascati models go down
to 0.6 \Msun. Moreover the SB99 models do not go beyond 1 Gyr.
Based on this comparison we adopt the SB99 fits for ages less than 700
Myrs and the results of the Frascati models for older ages.


\subsection{Uncertainties in the derived parameters}
\label{sec:5.4a}

The determination of the cluster parameters, age, mass and extinction,
by means of the 
two or thee dimensional maximum likelihood fitting method,
i.e. the 3(2)DEF-method, results not only in the values of 
$\log(\Mcl)$, $\log(t)$ and $E(B-V)$ but also in their maximum and
minimum acceptable values. 
The best-fit values are not necessarily in the middle of the minimum
and the maximum values. If we define
the uncertainties in the parameters, $\Delta E(B-V)$, $\Delta
\log(\Mcl)$ and $\Delta \log(t)$  
as half the difference between the maximum and mimimum values,
we find that
30 percent of the clusters have $\Delta E(B-V)<0.08$, 50 percent have
 $\Delta E(B-V)<0.11$ and 70 percent have $\Delta E(B-V)<0.15$.
For the uncertainties  in the mass determination we find 
$\Delta \log(\Mcl)<0.20$, $<0.33$ and $<0.55$ for 30, 50 and 70
percent of the clusters respectively. 
For the uncertainties  in the age determination we find 
$\Delta \log(t)<0.23$, $<0.39$ and $<0.69$ for 30, 50 and 70
percent of the clusters respectively. 
Taking the values for 50 percent of the clusters as representative,
we conclude that the uncertainties are $\Delta E(B-V)\simeq 0.11$,
$\Delta \log(\Mcl) \simeq 0.33$ and $\Delta \log(t) \simeq 0.39$.



\subsection{Contamination of the cluster sample by stars?}
\label{sec:5.5}

To estimate the number of stars that may contaminate our cluster
sample we use the stellar population in the solar neighbourhood.
>From the tabulated stellar densities as a function of spectral type
(Allen 1976)
we derived the number of stars brighter than a certain value of $M_V$
per pc$^3$. Using the mass density of 0.13 $\Msun$pc$^{-3}$, 
we derived the number of stars per unit
stellar mass. The results are listed in Table \ref{tbl:stars}.
The total stellar mass of the observed region of M51 
(see Fig. \ref{fig:sources}) is estimated
to be about $1/20$ of the total mass of $5~10^{10}$ \Msun\
in the disk of that galaxy (Athanassoula et al. 1987), i.e.
about $2.5~10^{9}$ \Msun. This implies that we can expect 
the following numbers of stars brighter than $M_V<-6.5$
per spectral type in the observed region:
4 (O), 13 (B), 6 (A), 6 (F), 6 (G), $<$1 (K) and $<$1 (M).
So we can expect of the order of 40 bright stars with $M_V<-6.5$
in the observed region of M51. However, these are all massive young
supergiants of which the vast majority will be in clusters! So the
number of bright stars outside clusters, that may contaminate our
sample of clusters will be considerable smaller, and we expect it to
be smaller than about 20 out of the total sample of 877. 
Moreover, we have shown above, from the lack of $O[III]$ emission, 
that the clusters in M51 have a shortage of massive stars with $M>30$
\Msun. We can expect this effect also to occur for the massive field stars.
This would reduce the the number of expected contaminating stars
even further.
Tests have shown that a considerable fraction of possibly remaining 
stellar sources will be eliminated by the
requirement that their energy distribution should fit that of cluster
models within a given accuracy. Based on all these considerations,
we conclude that contamination of our
cluster sample with very massive stars of $M_V<-6.5$ 
outside clusters is expected to be negligible.
 
\begin{table*}
\caption[]{Numbers of stars above a certain absolute magnitude limit
in the solar neighbourhood}
\begin{tabular}{c|cccccccc}
\hline\hline\noalign{\smallskip}
Limit & O & B & A & F & G & K & M & Total\\
\hline
\noalign{\smallskip}
$M_V < -5.5$ & -8.3 & -7.6 & -8.0 & -7.4 & -7.4 & -8.5 & -8.5 & -6.9 \\
$M_V < -6.5$ & -8.8 & -8.3 & -8.6 & -8.6 & -8.6 & $<$-9.5 & $<$-9.5 & -7.9\\
$M_V < -7.5$ & -9.8 & -9.3 & -9.6 & -9.6 & -9.6 & $<$-10.5 & $<$-10.5 & -8.9\\
\hline
\end{tabular}\\
(1) The data are in log (Number/ per \Msun)\\
(2) The number of stars with $M_V<-7.5$ was estimated to be about 10
times smaller than for $M_V<-6.5$ because of the small number of 
very massive stars formed, and their short evolution time.  
\label{tbl:stars}
\end{table*}


\section{The mass versus age distribution}
\label{sec:6}

We discuss the results from the fitting of the observed energy
distributions to those of cluster models with solar metallicity and with
and upper mass limit of 30 \Msun. We have checked that the 
fits with twice solar metallicity and with an upper limit of
100 \Msun\ give about the same results.

\begin{figure}
\centerline{\epsfig{figure=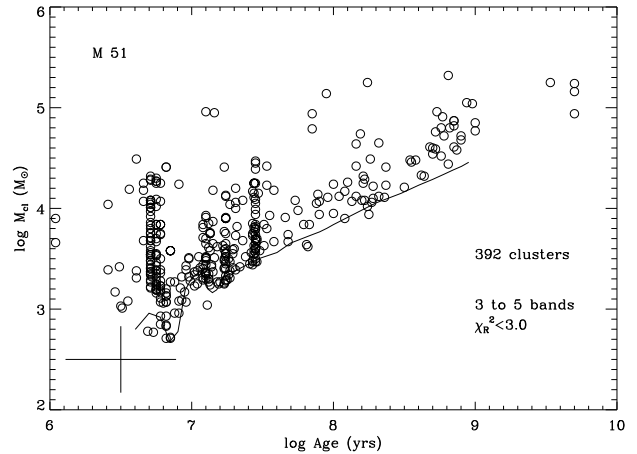,width=9.0cm}}
\caption[]{The mass versus age relation of the 392 clusters
 with $M_V < -7.5$,
whose energy distribution could be fitted with an accuracy of $\chinu<3.0$.
The cross shows the characteristic uncertainties  (Sect. 5.5).
The full line is the predicted fading line due to the evolution 
of clusters with \ebv=0 and with a limiting magnitude of 
$R_{\rm lim}=22.0$ 
at the distance of M51. This line roughly agrees with the observed
lower limit. 
The distribution is discussed in the text.}
\label{fig:mt}
\end{figure}

Fig. \ref{fig:mt} shows the mass versus age distribution of sources
with $M_v<-7.5$ (to eliminate possible stellar sources) and 
with energy distributions that could be fitted to that of a cluster
models with an accuracy of $\chinu \le 3.0$.
(The distributions for clusters with $\chinu \le 1$ or 10, not shown
here, show the same 
distribution, but with 294 and 508 clusters respectively.)
We see that the lower limit of the mass increases with increasing age,
from about 1000 \Msun\ at $\tau \simeq 5$ Myrs to about $5~10^4$ \Msun\
at 1 Gyr. This is due to the expected effect of fading 
of the clusters as they age (see Sect. 4.1).
The full line in the figure is the fading line in the $R$ magnitude
for clusters which have a limiting magnitude of $R_{\rm lim}=22.0$ at the
distance of M51 for a reddening of \ebv=0. 
The line in Fig. \ref{fig:mt} thus gives the
mass of the clusters, that reaches this magnitude detection 
limit, as a function of age.

These initial mass versus age distribution 
of the clusters in Fig. \ref{fig:mt} 
show the following characteristics:\\
-- (i) The lower mass limit increases with age due to the
    fading of the clusters. The observed lower limit agrees with the
    predicted ones for the $R$ band. There are even hints of the
    presence of the predicted dips in the lower limit near 
    $\log(t)$=6.8 and 7.2. This strengthens our confidence in the
    adopted models.\\
-- (ii) There are clear concentrations in the distribution at $\log(t)=6.70$
    and 7.45 and possibly also around $\log(t) \simeq 7.2$.
    These are due to the properties of the cluster models and
    the adopted method.
    In Fig. \ref{fig:sb99colours} we have shown that the colours of the
    models do not change monotonically with time, but that there are
    phases when the colours change rapidly with time. These phases occur
    in the range of $6.5 < \log(t) < 7.3$.
    Large changes in the colours of the models occur just
    before and after the peaks where the slopes of the curves in Fig. 
    \ref{fig:sb99colours} are large. 
    It is more difficult to fit the observed energy distributions
    with high accuracy to models in the age range when the spectral changes are
    large, than in the age range when the changes are small. So there is a
    tendency of the model fits to concentrate in agebins just 
    outside the age-regions of large spectral changes.
    This explains the concentrations and the voids in the derived age distributions
    of the clusters between about $6.5 < \log(t) < 7.5$.\\
-- (iii) The density of the points drops
    at $\log(t) > 7.5$. Since the age scale is logarithmic,
    we might have expected an {\it increasing} density of points
    towards the higher ages, which is not observed. 
    This is due to the disruption or dispersion of the clusters
    (see first paragraphs of Sect. \ref{sec:7}).
\\

We conclude that the mass versus age distribution agrees with the
expected evolutionary fading of the clusters and that the decrease
in numbers of clusters with age shows the affect of disruption/dispersion
of clusters with time. 
The concentrations of the clusters at ages
around $\log (t) = 6.8$, 7.2 and 7.45 are due to statistical effects
and do not represent periods of enhanced cluster formation.


\section{The cluster Initial Mass Function}
\label{sec:7}

The data in Fig. \ref{fig:mt} and the detailed study of this
distribution by Boutloukos \& Lamers (2002) show that the clusters 
in the inner spiral arms of M51 
disrupt on a time scale of about tens of Myrs. 
In fact, these authors derived the dependence of the disruption time on 
the initial mass of the clusters in the inner spiral arms of M51 as 

\begin{equation}
\log t_{\rm disr} = \log t_4 ~+~\gamma \times \log (M_{\rm cl}/10^4 \Msun)
\label{eq:tdis}
\end{equation}
with $\log t_4 = 7.64 \pm 0.22$ and $\gamma=0.62 \pm 0.06$ 
for the mass range of $3 \le \log (\Mcl / \Msun) \le 5.2 $,
where $M_{\rm cl}$ is the initial mass of the cluster.
We see that clusters with an initial mass larger than $10^4$ \Msun\
survive $4\times10^7$ years. Clusters with an initial mass of only
$10^3$ \Msun\ disrupt on a time scale of $1\times10^7$ yrs. 
This implies that {\it the cluster initial mass function  
cannot be derived from the total
sample of clusters, because the disruption will produce a strong bias
towards the more massive clusters}. However for the youngest clusters
with ages less than about 10 Myr  disruption is not yet an important
effect and, therefore,
these clusters can be used to derive the initial cluster mass
function.

\begin{figure}
\centerline{\epsfig{figure=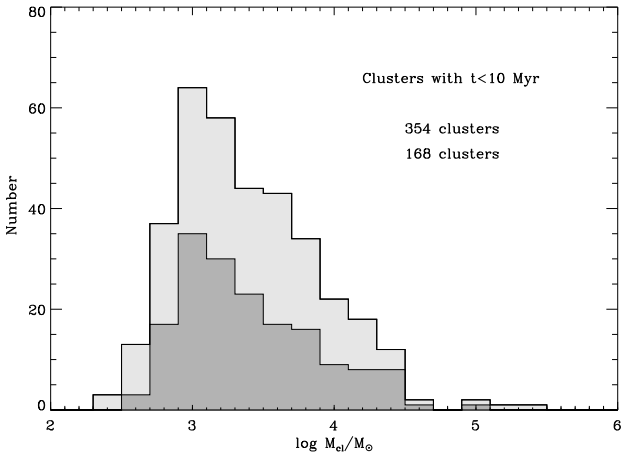,width=9.0cm}}
\caption[]{Histograms of the mass of clusters (in $\log M_{\rm cl}$)
with an age of 10 Myrs or younger. The light shaded area shows the 
distribution of all 354 clusters. The dark shaded area shows the
histogram of clusters with an accurate mass determination of
$\Delta log M_{\rm cl} < 0.25$.
The steep part at $\log(M) < 3.0$
is due to the detection limit. The decrease at $\log(M) >3.0 $  reflects
the slope of the cluster IMF.
(All masses have to be increased by a factor 2.1 if the lower limit of
the stellar mass is 0.2 \Msun\ rather than the value of 1 \Msun\
that was adopted in the Starburst99 models.)
}
\label{fig:imfhist}
\end{figure} 

Fig. \ref{fig:imfhist}
shows the resulting mass distribution of clusters with an age less 
or equal to 10 Myr for two samples of clusters.
The first sample contains all 354 clusters younger than 10 Myr.
The second sample contains 168 clusters in the same age range
but with a mass determination of $\Delta \log(\Mcl) \le 0.25$.
Both samples show the same characteristics:
a steep increase in number between $2.5 < \log \Mcl <3.0$ and a slow
decrease to higher masses. The steep increase is due to the detection
limit or the disruption of the low mass clusters. The slow decrease
reflects the cluster initial mass function (CIMF).
The decrease indicates that the clusters are formed with an CIMF
that has a negative slope of $d~\log (N)/d~\log (M)$, as expected.

\begin{figure}
\centerline{\epsfig{figure=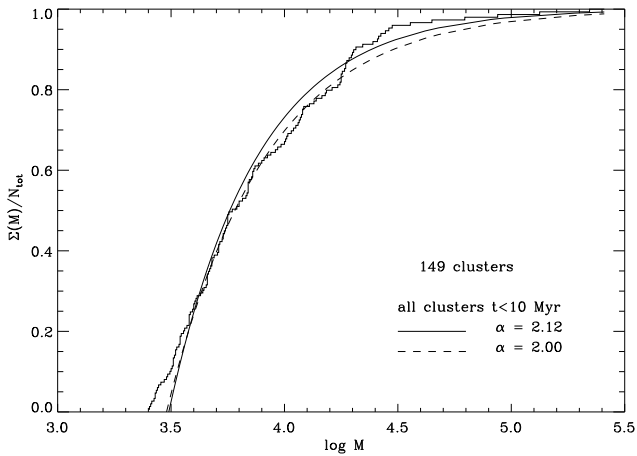,width=9.0cm}}
\caption[]{The cumulative mass distribution of 149 clusters with an
age less than 10 Myrs and an inital mass of $\log (M) > 3.40$.
The full line is the best power-law fits for an IMF 
with the value of $\alpha$ and \Mmin\ from Table \ref{tbl:imf}.
The dashed line is the fit with $\alpha=2.00$.}
\label{fig:cum}
\end{figure}

If the CIMF can be written as a power law of the type

\begin{equation}
N(M)~dM~ \sim ~ M^{\alpha}~dM~~~~~{\rm for}~~\Mmin<M<\Mmax
\label{eq:CIMF}
\end{equation}
then the normalized cumulative distribution will be 

\begin{eqnarray}
\Sigma (M)~&=&~A - B\times M^{-\alpha+1}~~~~~{\rm with} \nonumber \\
A~&=&~B \times \Mmin^{-\alpha+1} \nonumber \\
B~&=&~(M_{\rm min}^{-\alpha+1}~-~M_{\rm max}^{-\alpha+1})^{-1}
\label{eq:cum}
\end{eqnarray}
Fig. \ref{fig:cum} shows the observed and predicted normalized 
cumulative distribution of the 149 clusters with an age less
than 10 Myr  and with a mass $\Mcl >2.5~10^3~\Msun$ that are
 used for the determination of the CIMF.
We eliminated the clusters with $ \Mcl <2.5~10^3~\Msun$
from the sample because Fig. \ref{fig:imfhist} shows that the sample
may not be complete for smaller masses.
The cumulative distributions of the 84 clusters of $M>2.5~10^3~\Msun$,
younger than 10 Myr, which are fitted to the models with an accuracy
of $\chinu \le 3.0$, or of the 66 clusters with $\Delta \log(\Mcl) \le
0.25$, not shown here, (where $\Delta \log(\Mcl)$ is the uncertainty
in the mass determination),
have the same shape as the distribution in Fig.
\ref{fig:cum}.
For these three samples we have determined the values of $\alpha$ and
$M_{\rm min}$ by means of a linear regression under the reasonable assumption
that $\Mmin<<\Mmax$, which implies that $A \simeq 1$. The
resulting values of $\alpha$ and \Mmin\ are listed in Table
\ref{tbl:imf}.
The table shows that $\alpha \simeq 2.1$ for all three samples.

\begin{table}
\caption[]{The CIMF for clusters with $t<10$ Myr and
  $\Mcl < 2.5 \times 10^3 \Msun$}
\begin{tabular}{lrcc}
\hline\hline\noalign{\smallskip}
Sample  & Nr & $\log \Mmin$ & $\alpha$ \\
\hline\noalign{\smallskip}
All                       & 149 & 3.49 & 2.12 $\pm$ 0.26 \\
$\Delta \log \Mcl < 0.25$ &  66 & 3.48 & 2.04 $\pm$ 0.41 \\
$\chinu < 3.0$            &  82 & 3.46 & 2.16 $\pm$ 0.40 \\ 
\hline\noalign{\smallskip}
\end{tabular}
\label{tbl:imf}
\end{table}

The predicted cumulative distribution with the derived values of
$\alpha=2.12$ and \Mmin\ is shown in Fig. \ref{fig:cum} (full line).
The figure shows a slight underabundance of clusters in the range of
$4.0 < \log (\Mcl) < 4.3$ and a slight overabundance in the range of
$\log (\Mcl)>4.3$. In fact, for the mass range of $3.5 < \log (\Mcl) <
4.3$ a fit with $\alpha=2.00$, shown by dashed lines, fits the
distribution excellently.
We conclude that the CIMF of clusters younger than 10 Myr 
has a slope of $\alpha \simeq 2.1 \pm 0.3$ in the mass range of
$3.0 < \log (\Mcl) < 5.0$ and a slope of
$2.00 \pm 0.05$ in the range of $3.0 < \log(\Mcl) < 4.3~\Msun$.

The derived exponent of the cluster IMF is
very similar to the value of $\alpha = 2.0$ of young
clusters in the Antennae galaxies, as found by 
Zhang \& Fall (1999).  It shows that the IMF of clusters formed
in the process of galaxy-galaxy interaction is very similar
to the one of clusters formed in the spiral arms of a galaxy,
long after the interaction.
This mass distribution is also similar to that of
giant molecular clouds (e.g. McKee 1999; Myers 1999). This may support
the suggestion that the mass distribution of the clusters is
determined by the mass distribution of the clouds from which they originate.


\section{The cluster formation history} 
\label{sec:8}

One of the goals of this paper is to study the influence of the
interaction between M51 and its companion on the cluster formation in
the region of the inner spiral arms. 
To this purpose we compare the observed age distribution of the
clusters with predictions for a constant cluster formation rate.
This comparison is hampered by two effects: (a) the disruption of
clusters
and (b) the fading of clusters below the detection limit.
The disruption time of the clusters in the inner spiral arms of M51 is given
by Eq. (\ref{eq:tdis}).
We see that only clusters with an initial mass
larger than about $10^4$ \Msun\ will survive more than 40 Myrs.

Figure \ref{fig:agehist} shows the histogram of the formation
rates of the detected clusters, in number per Myr,
for  140 clusters with $\Mcl > 10^4$ \Msun\ and with an
energy distribution that is fitted with an accuracy of $\chinu \le 3.0$.
The sample of 111 clusters 
with $\Mcl > 10^4$ \Msun\ and with an age uncertainty of $\Delta
\log(t) < 0.25$,  and the full sample of 285 clusters
with $\Mcl > 10^4$ \Msun, not shown here, have a 
distribution very similar to the one shown in Fig. \ref{fig:cum}.

All three samples show about the same 
characteristics. \\
(a) There is a general trend of a decrease in the 
cluster formation rate towards increasing age. \\
(b) There is a steep drop around $\log(t)\simeq 7.5$.
 This drop is the  result of the concentration of clusters
at ages $\log(t)=7.45$ that was apparent in the mass versus age 
diagrams of Fig. \ref{fig:mt}, and was explained in Sect. 6.\\
(c) There is no clear evidence for a peak in the cluster formation
rate near $\log(t)\simeq 8.6$, which is the time of the interaction
of M51 with its companion, and which is the age of the huge
starburst in the nucleus of M51. There is a hint of a small peak 
around $\log(t) \simeq 8.1$. However this peak is only $2 \sigma$
high. Its reality has to be verified with a larger sample of clusters.\\
(d) The peak in the last bin of 5 Gyr contains all the clusters
ages older than 3 Gyr, because the 
cluster models that we used for fitting the energy distributions
do not go beyond 5 Gyr.\\
We have checked that these characteristics are not the result of the
binning process: the same features appear for different choices of the
binning parameters.\\

The general decrease in the formation rate of the observed clusters
is partly due to the evolutionary fading of the clusters and partly
due to the disruption of clusters.
In Sect. 6 and in Fig. {\ref{fig:mt} we have shown that clusters
with an initial mass of $10^4$ \Msun\ fade below the detection
limit when they are older than $\log(t)=8.2$. This can explain the
decrease in the formation rate of clusters older than about 150 Myrs.
However, clusters with $\Mcl>10^4~\Msun$ and younger than about
100 Myr should still all be detectable. The fact that we see a
decrease in the apparent cluster formation rate must thus be due to
the disruption of clusters (unless for some unknown reason the cluster
formation rate has been steadily increasing from 100 Myrs up to now, which
we consider unlikely).

It is interesting that the CFR in the inner spiral arms of M51, at a
distance of about 1 to 3 kpc from the nucleus, does not show any 
evidence for a peak
at the age of the strongest interaction of M51 with its companion
galaxy. The closest approach occurred about 250 - 400 Myrs ago
according to the dynamical models of Barnes (1998) or 400-500
and 50-100 Myrs ago according to Salo \& Laurikainen (2000). 
The only significant peak occurred at $\log(t) \simeq 7.4$. However we
attributed this peak to the large changes in the energy distributions
of the models in the age range of $6.5 < \log(t) < 7.5$, and not to a
real increase in the CFR.

\begin{figure}
\centerline{\epsfig{figure=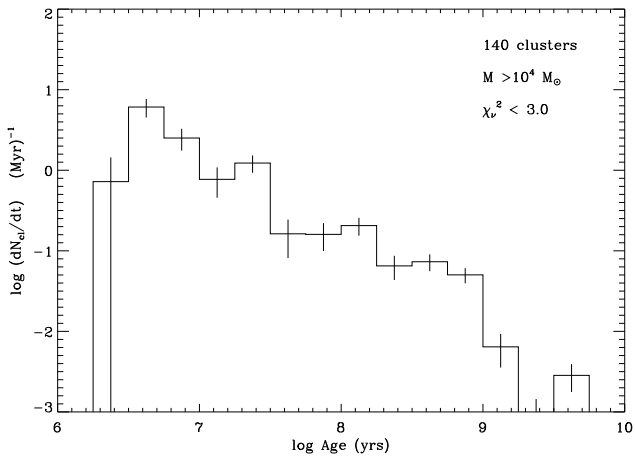,width=9.0cm}}
\caption[]{The formation rates of the observed clusters
 with an initial mass of $M_i > 10^4$ \Msun\, in number per Myr,
for clusters with an accuracy of the fit of the energy distribution of
$\chinu \le 3.0$.
The $ 1 \sigma$ uncertainty
due to the Poisson statistics in each bin ($\sigma = 1/\sqrt(N)$)
is indicated. The diagram shows that the formation rate of the
observed clusters decreases with age. This is due to the disruption of
the older clusters.}
\label{fig:agehist}
\end{figure}


\section{Summary and conclusions}
\label{sec:summary}

We have studied images of the inner spiral arms of the
interacting galaxy M51 obtained with the $HST-WFPC2$
camera in broad-band $UBVRI$  and in the narrow band
\halfa\ and \oiii\ filters. The study can be summarized as follows:

\begin{enumerate}
\item{We found a total  of 877 point-like objects, which are probably 
clusters. Many of the clusters are strong \halfa\ emitters, but none
of the clusters, not even the youngest ones, have an excess
of radiation in the \oiii\ line at 5007 \AA\ ($F502N$-filter). This 
suggests that the upper mass limit of the stars in the clusters
is about 25 to 30 \Msun.}

\item{We have compared their energy distributions with those of
Starburst99  cluster models (Leitherer et al. 1999) 
for instantaneous star formation 
with a stellar IMF of exponent 2.35, solar metallicity,
a lower and upper stellar mass limit of 1 \Msun\ and 
 30 \Msun\ respectively. 
The energy distributions were also
compared with those of the Frascati models (Romaniello 1998).
For clusters younger than 700 Myr the results from the fitting 
with the Starburst99 models were adopted because these models are more
accurate for young clusters and the fits of the energy distributions 
are better than those of the Frascati models. For older clusters
the results from the fits with the Frascati models were adopted.}

\item{For clusters that were observed in four or five bands 
a three dimensional maximum likelihood method was used
to derive the properties of the clusters from the comparison between the
observed and predicted energy distributions. The free parameters
are the age $t$ and $E(B-V)$, which together determined the
shape of the energy distribution, and the initial cluster mass
\Mcl\ which determines the absolute magnitude. For clusters that 
were not observed in all bands, the empirically derived lower magnitudes
limits were taken into account. }

\item{For clusters that were observed in only three bands the age and
mass were derived in a two-dimensional maximum likelihood fitting
of the energy distributions, with $t$ and \Mcl\ as free parameters.
The observed probability distribution of $E(B-V)$ was used as a
weighting factor in the fitting procedure.}
 
\item{The histogram of $E(B-V)$ is strongly peaked at very
small $E(B-V) \simeq 0$. All cluster have a reddening smaller than
$E(B-V)<1.0$ and 67 \% of the clusters have $E(B-V)<0.18$.}

\item{We have analysed the observed clusters also with
cluster models of higher metallicity, $Z=2\times Z_{\odot}$. These
higher metallicity models fit the observations considerably worse
than the solar metallicity models. For instance, for solar metallicity
models the energy distribution of 294 clusters 
can be fitted with an accuracy of $\chinu \le 1.0$ and 392 with
$\chinu \le 3.0$. For models with twice the solar metallicity these
numbers are respectively 138 and 217. So 
the energy distributions of the clusters support the adopted solar
metallicity.}

\item{The clusters have masses in the range of $2.5 < \log (\Mcl) <
    5.7$ and ages of $\log(t) >5.0$. 
These masses are the {\it initial} masses of the clusters, i.e. the
current mass corrected for stellar evolution effects, but not
corrected for evaporation or disruption.
All derived masses have to be multiplied by a factor 1.3 if the lower
mass of the stars is 0.6 \Msun, instead of the adopted 1 \Msun,
and by a factor 2.1 if the lower mass is 0.2 \Msun, as found for the
Orion Nebula cluster. }

\item{The distribution of the clusters in a mass-versus-age diagram
shows the predicted lower limit due to the evolutionary fading of the
clusters, including the dips at $\log (t)\simeq 6.8$ and 7.1. Three
apparent concentrations at $\log(t)=6.7$, 7.2 and 7.45 are not real but
due to the properties of the cluster models used.}

\item{About 60\% of the clusters are younger than 40 Myr. 
The number of older clusters is much less than expected for a  
constant cluster formation rate. This is partly due to the evolutionary
fading of low mass clusters below the detection limit, and partly due
to the disruption of the clusters. }

\item{ The cluster initial mass function (CIMF) was derived from the
cumulative mass distribution of clusters younger than 10 Myr, for
which disruption has not occured. The CIMF
has a slope of $\alpha = 2.1 \pm 0.3$ in the range of
$3.0 <\log (\Mcl) < 5.0$ and $\alpha= 2.00 \pm 0.05$ 
in the range of $3.0 < \log (\Mcl) < 4.5$ \Msun, 
for $N(\Mcl) \sim \Mcl^{-\alpha}$. This slope is the same to that
found in the interacting Antennae galaxies
(Zhang \& Fall 1999). 
Zhang and Fall deived a power law slope of the CIMF of $\alpha=1.95 \pm 0.03$
and $2.00 \pm 0.04$ for two cluster samples of the Antennae galaxies.
 The good agreement between these slopes and the one found by us 
suggests that 
$\alpha$ is about the same for cluster formation triggered by 
strong galaxy-galaxy interactions, such as presently going on in 
the Antennae, as for cluster formation that is not dominated by the interactions.}

\item{The age distribution of clusters with $\Mcl > 10^4$ \Msun,
is used to derive the history of the cluster formation rate (CFR). 
There is a general
trend of a decrease of the formation rate of the observed clusters
with age. It is unlikely that the real CFR has
been increasing continuously from about 1 Gyr to the present time.
The decrease of the CFR with age of clusters younger than about 100 Myr 
cannot be due to evolutionary fading, but it is due to the disruption
of clusters. For clusters older than 200 Myr the decrease of the
derived CFR could, at least partly, be due to evolutionary fading.
}

\item{There is no evidence for a peak in the CFR at about 400 Myr,
    which
is the time of the interaction of M51 with its companion and the age
of the huge starburst in the nucleus.}

\end{enumerate}

In a forthcoming paper we describe 
the cluster formation as a function of location 
in a large part of M51, using the same methods as used here 
(Bastian et al. 2002).
The disruption of clusters in M51, derived from the results of the
study presented here, are described by Boutloukos \& Lamers (2002).

\section{Acknowledgement}
H.J.G.L.M.L. and N.B. are grateful to the Space Telescope Scence Institute
for hospitality and financial support during several stays.
We thank Claus Leitherer for help and advice in the calculation
of the cluster models.
Support for the SINS program GO-9114 was provided by NASA through a
grant from the Space Telescope Science Institute, which is operated by
the Association of Universities for Research in Astronomy, Inc. under
NASA contract NAS 5-26555.
N.B. ackowledges a grant from the Netherlands Organization for
Scientific Research. We thank the unknown referee for constructive
comments that resulted in an improvement of this paper.




\end{document}